\documentclass{vgtc}                          % final (conference style)
\ifpdf%                                % if we use pdflatex
  \pdfoutput=1\relax                   % create PDFs from pdfLaTeX
  \usepackage{graphicx}                % allow us to embed graphics files
  \DeclareGraphicsExtensions{.pdf,.png,.jpg,.jpeg} % for pdflatex we expect .pdf, .png, or .jpg files
\else%                                 % else we use pure latex
  \usepackage{graphicx}                % allow us to embed graphics files
  \DeclareGraphicsExtensions{.eps}     % for pure latex we expect eps files
\fi%

\graphicspath{{figures/}{pictures/}{images/}{./}} % where to search for the images

\usepackage{microtype}                 % use micro-typography (slightly more compact, better to read)
\PassOptionsToPackage{warn}{textcomp}  % to address font issues with \textrightarrow
\usepackage{textcomp}                  % use better special symbols
\usepackage{mathptmx}                  % use matching math font
\usepackage{times}                     % we use Times as the main font
\usepackage{cite}                      % needed to automatically sort the references
\usepackage{booktabs}                  % only used for the table example
\usepackage[hyphens]{url}
\usepackage{amsmath}
\usepackage{enumitem}
    \setlist{nosep}

\onlineid{0}

\vgtccategory{Research}

\title{HeaRT: Health Record Timeliner to visualise patients' medical history from health record text}

\author{Shuntaro Yada\thanks{e-mail: s-yada@is.naist.jp}\\ %
        \scriptsize Nara Institute of Science and Technology %
\and Eiji Aramaki\thanks{e-mail: aramaki@is.naist.jp}\\ %
     \scriptsize Nara Institute of Science and Technology %
}

\abstract{%
Electronic health records (EHRs), which contain patients' medical histories, tend to be written in freely formatted (unstructured) text because they are complicated by their nature. 
Quickly understanding a patient's history is challenging and critical because writing styles vary among doctors, which may even cause clinical incidents. 
This paper proposes a Health Record Timeliner system (HeaRT), which visualises patients' clinical histories directly from natural language text in EHRs. 
Unlike only a few previous attempts, our system achieved feasible and practical performance for the first time, by integrating a state-of-the-art language model that recognises clinical entities (e.g.\ diseases, medicines, and time expressions) and their temporal relations from the raw text in EHRs and radiology reports. 
By chronologically aligning the clinical entities to the clinical events extracted from a medical report, this web-based system visualises them in a Gantt chart-like format. 
Our novel evaluation method showed that the proposed system successfully generated coherent timelines from the two sets of radiology reports describing the same CT scan but written by different radiologists.
Real-world assessments are planned to improve the remaining issues.%
} % end of abstract

\CCScatlist{
  \CCScatTwelve{Human-centered computing}{Visu\-al\-iza\-tion systems and tools}{}{};
  \CCScatTwelve{Applied computing}{Health informatics}{}{};
  \CCScatTwelve{Computing methodologies}{Information extraction}{}{};
}

\begin{document}

%% The ``\maketitle'' command must be the first command after the
%% ``\begin{document}'' command. It prepares and prints the title block.

%% the only exception to this rule is the \firstsection command
% \firstsection{Introduction}

\maketitle

\section{Introduction}

Patients' clinical histories are typically managed through electronic health records (EHRs), which often contain a substantial amount of freely formatted (unstructured) text.
Unlike structured data, such as numerical or categorical inputs about medication and tests, medical information described in unstructured text is not easily retrievable from straight-forward aggregation systems for EHRs, although such text explains detailed states of patients at a higher resolution than structured inputs.
Natural language processing (NLP) techniques make such text machine-readable.
A popular application is the combination of named entity recognition (NER) and relation extraction (RE), which identifies medical terms and concepts written in the text and resolves the relations between them \cite{Xue2019-bu,Li2018-fy,Christopoulou2020-dy}. 
Such NLP solutions contribute to (a) clinical big data analysis using all available EHRs of many patients and (b) personalised medicine using the EHRs of a patient.
For both directions, timeline-based visualisation of clinical history is known as to be useful \cite{Rind2013-yc}.

Although some graphical applications for this purpose have previously been proposed, almost all systems only expect explicitly structured inputs in EHRs and ignore unstructured textual inputs therein \cite{Rind2013-yc,Roque2010-bi}.
Moreover, only a few studies have developed preliminary prototypes to visualise patients' timelines from text \cite{Jung2011-me,Sultanum2018-pj}, although substantial work has addressed component-level sub-tasks regarding temporal
reasoning from text \cite{Sun2013-fh,Olex2021-wi}. 
Besides the technical difficulty, the lack of practical timeline visualisers may also stem from the fact that no overall goal has been determined and shared among relevant communities.

% \begin{figure}[t]
%     \centering
%     \includegraphics[width=\linewidth]{figures/heart-screen.png}
%     \caption{Example of HeaRT system output visualising a radiology report's timeline (labels are written in Japanese). 
%     Horizontal bars in each row denote clinical facts described in the report.}
%     \label{fig:heart_sample}
% \end{figure}

\begin{figure*}[tbp]
    \centering
    \includegraphics[width=\linewidth]{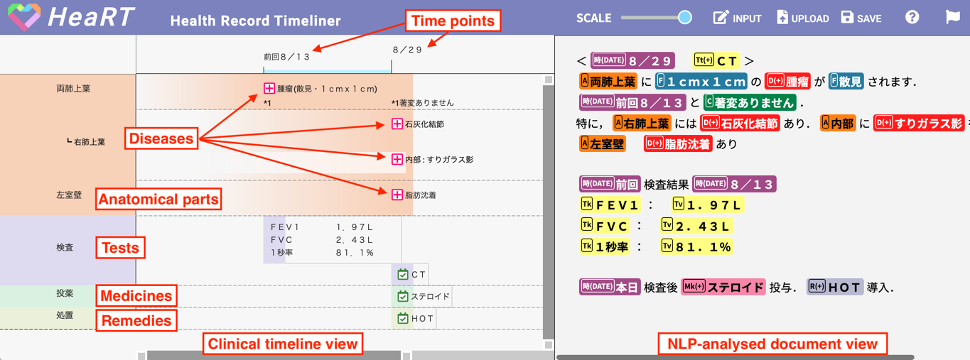}
    \caption{An example timeline visualising a Japanese health record using our system HeaRT.
    The input record is shown on the right side, annotated with medical entities such as diseases, medicines, time expressions.
    On the left side, the horizontal bars in each row denote clinical facts described in the report, which are chronologically aligned to time expressions found also in the report.}
    \label{fig:heart_sample_doc}
\end{figure*}

Thus, we propose a patient-timeline visualisation system named \textbf{\textsc{Hea}lth Record Timeliner} (\textbf{HeaRT}), which is capable of analysing unstructured free text in EHRs by using the state-of-the-art NLP techniques (Figure~\ref{fig:heart_sample_doc}). 
As a starting point, we first focus on the use-case for Contribution (b), i.e.\ visualising the medical history of a patient.
This would allow doctors to grasp the patient's status more quickly and precisely than manually reading or skimming their EHRs.
Such a patient-wise medical timeline visualisation works like Google's search-result display, where a summary of each web page is provided through a short ``snippet'' of its contents.
% Similarly, a quick summarising technique to represent a patient history would help EHR database systems.
We believe that this application can contribute to daily medical examinations in many hospitals.
The possible practical use cases are as follows:

\begin{itemize}
\item
  Understanding long-term hospitalised patients speedily (for the
  doctors taking over a case)
\item
  (Co-)medical staff education (for students or fresh workers to grasp a case
  more easily)
\item
  Patient-doctor communication (to enhance informed
  consent)
\end{itemize}

Our proposed HeaRT system consists of an NLP module (backend) to analyse EHR text and a user-interface module (frontend) to visualise the patient's clinical timeline.
We demonstrate its functionality on manually annotated Japanese clinical documents.
A demo is available at \url{https://aoi.naist.jp/prism-heart/}. 

\section{Related Work}

Timeline visualisation of health records itself is not a new idea.
Rind et al.~\cite{Rind2013-yc}, which provided an overview of visualisation systems for EHRs, reported that four applications which shows clinical events recorded in EHRs had been proposed as of 2013.
These applications visualise clinical events such as tests and treatments from pre-formatted data, i.e.\ tuples of event category, value, and date \cite{Plaisant1996-sc,Combi1999-im,Bui2007-ps}.
Although EHR systems may allow doctors to input clinical events in such a structured format, a substantial amount of important information has still been written in text areas that accept unstructured text.
However, generating timeline visualisation from text (i.e. text-to-timeline) in medical and clinical domains has been investigated only a few times.
An earliest study of text-to-timeline systems for EHRs \cite{Hallett2008-df} proposed some aesthetic requirements for visualising clinical history of a patient, although any actual NLP methodology to achieve such visualisation was not considered or discussed.
Jung et al.~\cite{Jung2011-me} proposed a text-to-timeline system that directly processes EHR text using classical NLP techniques (e.g.\ tokenisation, NER, and dependency parsing) and handwritten rules of temporal reasoning. 
This study, nonetheless, evaluated the system with only one particular sample of EHRs.
To adapt this system for different clinical texts, temporal reasoning rules must be updated manually.
Besides, the output visualisation proposed in the study borrowed a general-purpose timeline visualiser, which was not optimised for clinical information that comprises complex categories and conditions.
As concluded in a recent review of temporal reasoning research in clinical texts~\cite{Olex2021-wi}, no end-to-end practical system that visualise patients' medical history from natural language text in EHRs has ever be developed.

The difficulty in medical/clinical text-to-timeline stemmed from the performance of temporal information extraction using NLP~\cite{Sun2013-fh,Olex2021-wi}.
This is largely because temporal information is mentioned in implicit ways, such as ``past four days'', ``1 month after surgery'', and ``early in the morning''.
To build a timeline, such phrases should first be extracted as time representations, and then normalised into the exact date and time in order to sort these time points.
Clinical entities found in text, then, need to be allocated to these time points (i.e.\ temporal relation identification).
Recent advances in NLP achieved practically high performance even in such temporal tasks by using large pretrained language models such as BERT~\cite{Devlin2019-hz}.
We adapted BERT for precise temporal reasoning in clinical domains, which allowed us to develop a text-to-timeline system that interprets various types of medical/clinical documents with practical performance.

% \section{Methods}

\section{HeaRT System Architecture}

\begin{figure}[tbp]
    \centering
    \includegraphics[width=\linewidth]{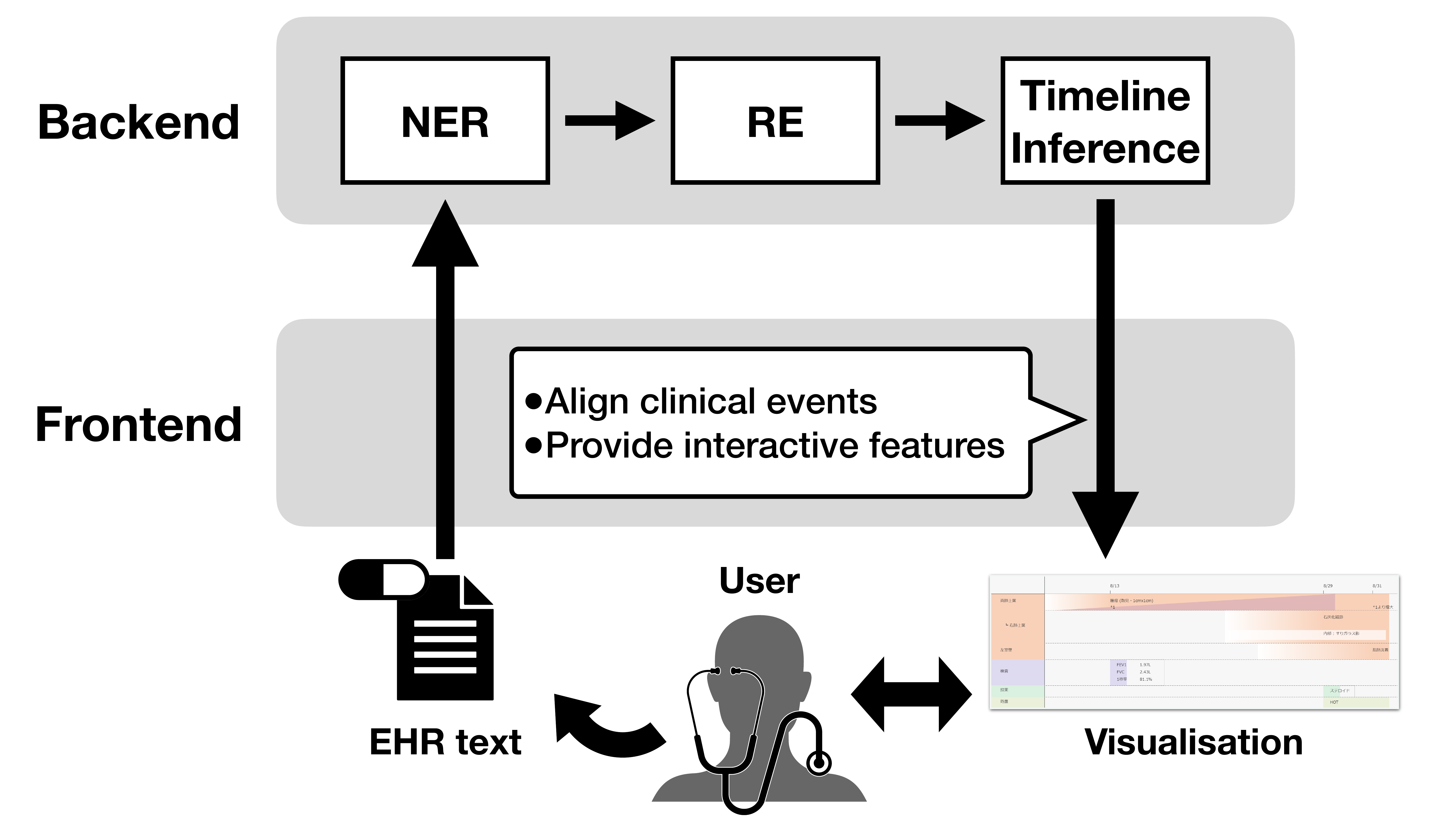}
    \caption{System architecture of HeaRT}
    \label{fig:architecture}
\end{figure}

The proposed system consists of two modules: a backend NLP pipeline and
a frontend visualisation presenter (Figure~\ref{fig:architecture}). 
The frontend module takes an EHR document (raw text data), and passes it to the backend module which identifies clinical named entities and the relations among them by using a machine learning-based NLP model.
Next, the named entities, as clinical events of the patient, are associated with the time points described in the document based on the relations among them.
The chronologically aligned named entities are then sent back to the frontend, which finally draws a timeline visualisation of the document.

\subsection{Backend}

\subsubsection{NER and RE model: JaMIE}
\label{subs:nlp_models}

The backend module provides a machine learning model that solves NER and RE simultaneously.
We adopted JaMIE~\cite{Cheng2022-lq}, which is based on BERT~\cite{Devlin2019-hz} and was trained on two types of Japanese clinical corpora: health records and radiology reports.
% These corpora target critical lung diseases, which we believe does not violate the generalisability of our system.
% In particular, for NER, we adopted Yada et al. 's model implementation~\cite{Yada2020LREC}, whereas the RE model, on the other hand, is an adaptation of a dependency parsing model \cite{zhang-etal-2017-dependency-parsing} that labels relations between pairs of named entities in the input string.
% \paragraph{Named Entity and Relation Representations}
% Both clinical corpora for training NLP models (written in Japanese) 
Both corpora were annotated with medical entities and relations according to an annotation scheme proposed by Yada et al.~\cite{Yada2020LREC,Yada2021-vr}, and thus the model can annotate unseen clinical texts with following the scheme.
This model and annotation scheme achieve the broad coverage of medical entity types and relations found in typical clinical text, including time expressions and relations, which are crucial for our visualisation.
Also, JaMIE achieves a reasonably high performance, i.e., 0.85--96 and 0.71--87 F1 scores in NER and in RE.

The entity types are as follows: 
% \begin{itemize}
    % \item 
        \emph{Disease} (and symptoms), 
    % \item 
        \emph{Anatomical} (parts), 
    % \item 
        \emph{Feature} (to modify diseases and anatomical parts), 
    % \item 
        \emph{Change} (of diseases and anatomical parts),
    % \item 
        \emph{Time} (expressions, i.e. TIMEX3 \cite{Pustejovsky2003-ui}), 
    % \item 
        \emph{Test}, 
    % \item 
        \emph{Medicine}, 
    % \item 
        \emph{Remedy}, and 
    % \item 
        Clinical Context (\emph{CC}). 
% \end{itemize}
The Test and Medicine entities include key-value pairs (e.g.\ a medicine name and its dosage value).
Disease entities have a \texttt{certainty} attribute, representing whether the disease was confirmed in the patient or not (i.e.\ \textit{positive}, \textit{negative}, \textit{suspicious}, or \textit{general} mention).
TIMEX3 entities adopts a \texttt{type} attribute to distinguish time representations into: \textit{date}, \textit{time}, \textit{duration}, \textit{set}, \textit{age}, \textit{medical}, and \textit{misc}.
Test, Medicine, and Remedy entities have a \texttt{state} attribute to describe whether these treatments were executed or not (i.e.\ \textit{executed}, \textit{negated}, \textit{scheduled}, and in \textit{other} states).

To encode clinically important information, the following medical relations are defined between two medical entities: 
% To visualise medical named entities in a meaningful timeline, we defined \emph{basic relations} and \emph{time relations}. 
% The basic relations, which involve clinical semantics, include the following relation types: 
% \begin{itemize}
    % \item 
        \emph{changeSbj} (from a Change entity to the changed subject),
    % \item 
        \emph{changeRef} (from a Change entity to the referenced entity),
    % \item 
        \emph{featureSbj} (from a Feature entity to the modified entity),
    % \item 
        \emph{subRegion} (from an Anatomical/Disease entity to the contained entity), and 
    % \item 
        \emph{keyValue} (from a Medicine/Test-key entity to the corresponding Medicine/Test-value entity).
% \end{itemize}

Furthermore, time relations from an entity (E, including TIMEX3) to a TIMEX3 entity (T) are defined as follows: 
% \begin{itemize}
    % \item 
        \emph{timeOn} (when E happened at T), 
    % \item 
        \emph{timeBefore} (when E happened before T), 
    % \item 
        \emph{timeAfter} (when E happened after T), 
    % \item 
        \emph{timeBegin} (when E started at T), and 
    % \item 
        \emph{timeEnd} (when E finished at T). 
% \end{itemize}
% In particular, for the time relations, we adapted 
T includes document creation time (DCT) to determine the relative chronological relations among other Ts.
% Thanks to these annotations, we can interpret precise medical information chronologically from clinical text.

\begin{figure*}[tbp]
    \centering
    \includegraphics[width=\linewidth]{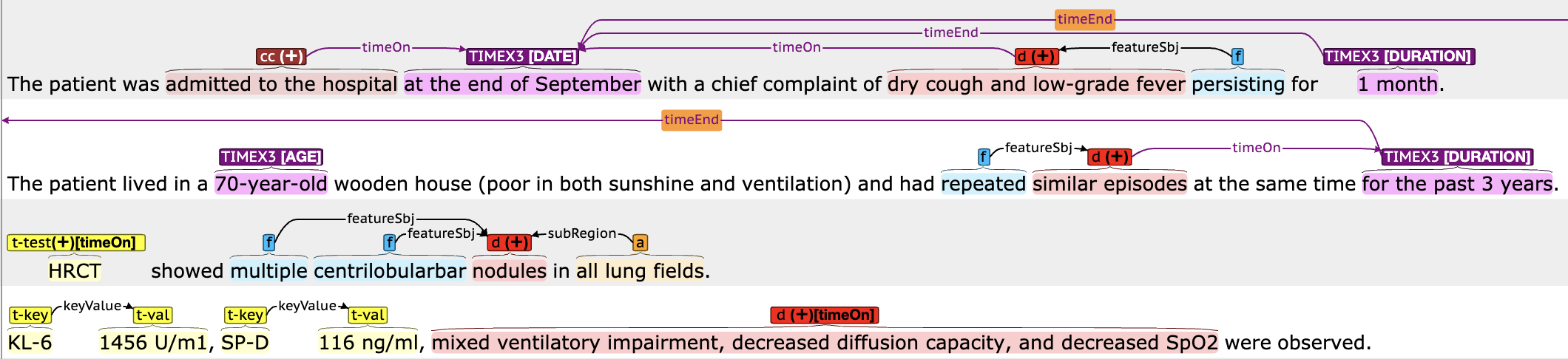}
    \caption{A sample English snippet of EHRs annotated with clinical named entities and relations}
    \label{fig:anno_eg}
\end{figure*}

Figure~\ref{fig:anno_eg} shows an English example of the annotation for non-Japanese readers; note that the model works in Japanese.

\subsubsection{Timeline Inference}

To chronologically organise medical entities obtained from JaMIE, the timeline inference module processes them as follows:

\begin{enumerate}
    \item  Group entities into time clusters, which are bags of the entities that belong to the same time point or duration, based on the time relation ``timeOn''.
    \begin{itemize}
        \item Disease and Anatomical entities that contain other entities via the subRegion relation are bundled into single entities.
        \item Feature and Change entities are included as supplemental information into the entities to which they are attached by the featureSbj and changeSbj relations.
    \end{itemize}
    \item Sort the time clusters in the chronological order based on the surface temporal expression of TIMEX3 entities, as well as time relations among the clusters using topological ordering. %\cite{Kahn1962-eu}
    \item Infer the time spans of the entities that started at a certain time point and ended at another time point based on the time relations annotated to the entities.
\end{enumerate}

These procedures are based on the idea of narrative containers \cite{pustejovsky-stubbs-2011-increasing,Styler2014}.
Finally, chronologically bundled and ordered medical entities are sent back to
the frontend in a JSON format.
A reference implementation is published at: \url{https://github.com/shuntaroy/prism-relanno}.

\subsection{Frontend}

% \begin{figure}[tbp]
%     \centering
%     \includegraphics[width=\linewidth]{figures/heart-screen-with-doc.png}
%     \caption{An example visualisation of a Japanese EHR using HeaRT}
%     \label{fig:heart_sample_doc}
% \end{figure}

% Figure~\ref{fig:heart_sample_doc} shows an example display from an EHR text, produced by the frontend module.
% We elaborate on the methods and features in the sections below.

\subsubsection{Entity Alignment}

Following the direction of prior studies \cite{Rind2013-yc}, we adopted a Gantt chart-based format, which aligns events that span in a horizontal row grouped into multiple rows of categories.
Our format consists of a chronological axis (horizontal) and entity-type axis (vertical). 
Given the JSON data created by the timeline inference module, the frontend aligns the medical entities for each time cluster according to the following rules:
\begin{enumerate}
    \item In Disease rows (orange-coloured), which are divided into the parent Anatomical parts, put the Disease entities to the corresponding Anatomical rows to which they belong
    \begin{itemize}
        \item If a disease entity contains another disease, it is placed immediately under the disease entity.
        \item Supplemental information such as Feature and Change entities is placed in the right margin of the entity.
        \item Diseases that do not belong to particular Anatomical parts are organised into the ``Diseases'' row (pink-coloured).
    \end{itemize}
    \item For Test (violet-coloured) and Medicine (green-coloured), combine corresponding key-value pairs into a table structure and put them in the time span to which these entities belong.
    \item Remedy and CC entities are assigned to a discrete row called ``clinical treatment'' (light green-coloured).
\end{enumerate}

\subsubsection{Interactive Features}

The major interactive features are as follows:
\begin{itemize}
    \item Change the entire width of the timeline by using a slider in the header, so that the users can choose from an overview mode to a detailed-view mode.
    \item Show the NER-ed document side by side by pressing a flag icon in the header.
    \begin{itemize}
        \item Highlight the corresponding entities by hovering the mouse over between the timeline view and the document view, so that the users can check the detailed nuances of the original text.
    \end{itemize}
    \item Save the timeline as a PNG image by pressing the SAVE button in the header.
    \item Input an EHR document to visualise by pressing the INPUT button, which opens a text-input dialogue.
\end{itemize}
We released a reference implementation at: \url{https://github.com/sociocom/prism-heart}.

\section{Results}

To evaluate our HeaRT system, we assessed the ``accuracy'' of the frontend using the human-annotated (gold) documents.
We introduce two methodologies to check the output visualisation quality: \textit{coherence} check and \textit{precision} check.
The former measures the robustness to the input text and the latter assesses the accuracy of the visualised output.
% We used the human-annotated (gold) documents. %for this visualisation evaluation. 
here we report a part of the evaluation results, the full detail of which is available at: \url{https://github.com/shuntaroy/heart-evaluation}.

\subsection{Coherence check}
\label{visualisation-coherence-check}

\begin{figure*}[tbp]
    \centering
    \includegraphics[width=\linewidth]{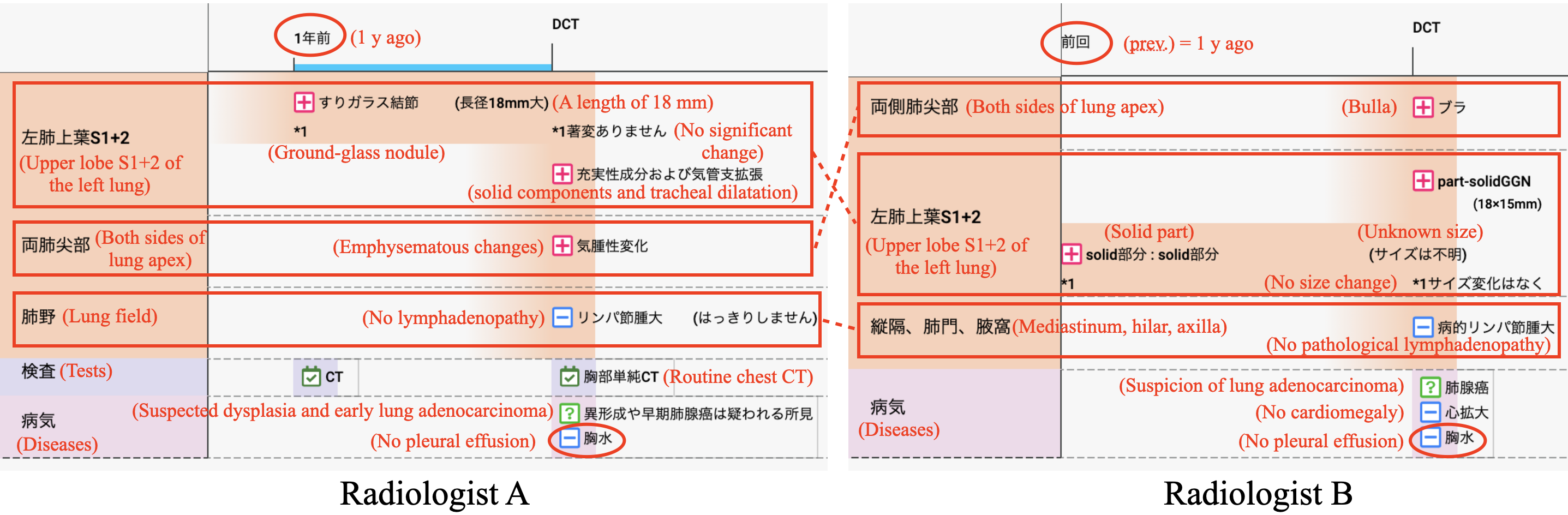}
    \caption{Timeline visualisations of two radiology reports for an identical CT image  independently written by two radiologists}
    \label{fig:coherence_check}
\end{figure*}

% To the best of our knowledge, the evaluation of EHR visualisation is a novel challenge.
We carried out a visual matching evaluation using a ``comparable'' clinical corpus which consists of multiple radiology reports where different radiologists independently describe identical CT scan images \cite{Nakamura2022-ty}.
We assume that a coherent visualisation system can produce timelines similar to these reports regardless of the difference in surface textual expressions.
From the corpora, we selected two radiology images of one-year-later follow-up of a lung nodule \cite{Weerakkody2015-ba,Bell2018-at}, and thus two reports for each (four in total) were fed into HeaRT.
% One image  was taken 
%     from a 74-years-old woman 
%     without episodes.
%     % as one-year-later follow-up of a lung nodule.
% The other image \cite{Bell2018-at} was taken
    % from a 50-years-old woman
    % without episodes.
    % as one-year-later follow-up of a lung nodule, and
    % the entire length of the pathological change is 12 mm, which was increased from 9 mm one year ago.
% Figure~\ref{fig:coherence_check2} displays the result.
% These four radiology reports were manually labelled with the JaMIE compatible annotation. 
We report the output of one image \cite{Weerakkody2015-ba} in Figure~\ref{fig:coherence_check}.
The result of the other image is reported in the evaluation repository above.
% and \ref{fig:coherence_check2}

% \begin{figure}[tbp]
%     \centering
%     \includegraphics[width=\linewidth]{figures/coherence-check2.png}
%     \caption{Timeline visualisations of two radiology reports for an identical CT image independently written by two radiologists}
%     \label{fig:coherence_check2}
% \end{figure}

% Although our input was manually annotated in advance, machine learning models can automatically label the data \cite{Yada2020LREC}. If our visualisation app is concatenated with such models, a large-scale deployment of timeline visualisation becomes possible within hospitals.

\subsection{Precision check}
\label{visualisation-precision-check}
% 小林先生の評価

% As a more direct and standard method, we evaluated the precision of visualisation by 
To check whether the visualisations correctly interpret and represent the input documents, we asked a doctor to compare input clinical documents and their output visualisations.
For all entities visualised by the system, we asked the doctor to assess the following points: whether the entities are located at the correct time point (\textit{OnSet}), whether the entities span the correct duration when they exist or are applied (\textit{Duration}), and whether the changes happening to diseases are described correctly (\textit{ChangeInfo}). 
We used each one case report and radiography report for this evaluation.
% ; the translated snippets of which are available in the Appendix.
Table~\ref{tab:precision_check} reports the aggregated accuracy, whereas the visualised timelines and document-wise accuracy results are available in the aforementioned evaluation repository.

\begin{table}[tbp]
    \caption{Accuracy values for each clinical entity to be placed at precise positions in the generated visualisations of different clinical documents}
    \label{tab:precision_check}
    \centering
    \small
    \begin{tabular}{lrrr}
    \toprule
        {} & OnSet & Duration & ChangeInfo \\
    \midrule
        Case report & 18/20 (90.0\%) & 18/20 (90.0\%) & --- \\
        Radiology report & 15/17 (88.2\%) & 15/15 (100\%) & 1/2 (50.0\%) \\
    \bottomrule
    \end{tabular}
\end{table}

\section{Discussion}

% NLP results
% The radiology reports' performance suggests that by concentrating annotation efforts on a specific report type the system achieves high F1 with sufficient training data. Compared to 95.30 NER F1 reported by~\cite{Yada2020LREC}, our NER score outperforms their score by 0.35 with the additional CRF layer. The RE model obtains 86.53 F1 of the radiography interpretation reports and 71.04 F1 of the medical history reports. 
% First, we review the NLP model evaluation.
% W observed high performance throughout each pipeline stage both in radiology reports and health records.
% The scores for radiology reports outperformed those for health records probably because radiology reports are written in a simpler format than health records.
% However, we still obtained more than 0.71 F1-scores in each pipeline stage for the health records processing, meaning that our NLP pipeline produces practical-enough predictions for different types of clinical text.
% Thanks to the nature of BERT we adopted as the Sentence Encoder, these scores can further be improved by increasing the training data size and model parameter size \cite{Devlin2019-hz}.

% Next, we discuss the visualisation evaluation.
In the coherence check, both timelines successfully displayed the key finding about the part-solid GGN in the left lung's S1+2 area, which is an expected common interpretation of the given image.
Other findings in the different areas are also identical between the two doctors, while textual surface expressions differ somewhat, as shown in the figure.
In fact, only 12.6\% of the word bi-grams are shared between the two reports.
Despite this diversity, the resulting timeline figures look similar, which demonstrates the degree of coherence of the system output.
In practical use cases, however, these differences in textual surface expressions should be mapped to medically standardised names.
We plan to include clinical entity linking or normalisation techniques to
solve this issue.

In the precision check, both types of the clinical documents were processed with high accuracy.
Most errors were derived from the bugs in the user-interface module, such as failure to show ``cancelled'' icons for Test and Medicine entities.
This issue may well happen because the timelines are drawn by a complex and dynamic procedure, but can be reasonably fixed by software engineering.
The doctor who conducted the precision check additionally noted repetitive occurrences of the identical clinical entities.
A clinical concept can be mentioned multiple times both in the identical expression and in different expressions.
We also plan to add coreference resolution to our NLP backend module in order to tackle this problem.

We should mention an intrinsic limitation of our evaluation scheme, i.e. the lack of discussion about which information in the clinical text must be visualised in the timeline.
To assess this ``recall''-oriented performance, we plan to cooperate with hospitals and doctors; we will ask actual doctors and co-medical staff to use HeaRT for real-world EHR text so that they can give their own practical feedback.
% We started by calling for relevant doctors for this evaluation.

In this research, Japanese clinical text was targeted. 
Since our backend NLP model adopts state-of-the-art machine learning techniques, our system can be applied to other languages without changing the frontend user interface and the timeline inference module in the backend.
For instance, creating annotated corpora in English enables HeaRT to visualise English EHRs.

\section{Conclusions}

We developed the HeaRT system, which visualises the clinical history of a patient by interpreting free text contained in EHRs.
The proposed architecture consisting of 
    (a) NLP models that directly analyse such unstructured text to extract and sort medical information chronologically, and 
    (b) a user interface that aligns it in a concise timeline view.
Unlike most of the existing clinical timeline visualisers that use structured inputs only \cite{Rind2013-yc,Roque2010-bi}, our novel graphical application immediately allows any doctors to try out our system by inputting their own EHR text regardless of which EHR systems they use.
Using the state-of-the-art NLP and a simply organised visualisation format is our improvement from the former text-to-timeline applications \cite{Jung2011-me,Sultanum2018-pj}. 

While we first focused on visualising single patients, one future step is to extend the HeaRT system to process multiple patients' histories at the same time. 
This may help in clinical big data analysis and precision medicine more.

%% if specified like this the section will be committed in review mode
\acknowledgments{
This work was supported by MHLW Program Grant Number JPMH21AC500111, Japan.
    % This work was supported by Cabinet Office, Government of Japan, Public/Private R\&D Investment Strategic Expansion Program (PRISM).
    % Radiology reports used in Result section are provided by Yuta Nakamura, MD at University of Tokyo.
}

\bibliographystyle{abbrv-doi}

\bibliography{heart}
\end{document}